\documentclass[sigconf, nonacm]{acmart}

\newcommand\vldbdoi{XX.XX/XXX.XX}

\newcommand\vldbavailabilityurl{}

\usepackage{algorithm}
\usepackage{balance}
\usepackage[noend]{algpseudocode}
\usepackage{caption}
\usepackage{subcaption}
\usepackage{xcolor}
\usepackage{varwidth}
\usepackage{mdwlist}
\usepackage{tabularx}
\usepackage{tikz}
\usetikzlibrary{positioning}
\usepackage{xspace}
\usepackage{url}
\usepackage{titlesec}
\usepackage{subcaption}
\usepackage{tabularx}
\usepackage{array,graphicx}
\usepackage{booktabs}
\usepackage{pifont}
\usepackage{xspace}

\newcommand*\OK{\ding{51}}
\newcommand*\circled[1]{\tikz[baseline=(char.base)]{
		\node[shape=circle,draw,inner sep=1pt] (char) {#1};}}

\begin{document}
\title{Zero-Shot Cost Models for \\Out-of-the-box Learned Cost Prediction}

\author{Benjamin Hilprecht}
\affiliation{
  \institution{TU Darmstadt}
}

\author{Carsten Binnig}
\affiliation{
	\institution{TU Darmstadt}
}

\begin{abstract}
In this paper, we introduce zero-shot cost models which enable learned cost estimation that generalizes to unseen databases. In contrast to state-of-the-art workload-driven approaches which require to execute a large set of training queries on every new database, zero-shot cost models thus allow to instantiate a learned cost model out-of-the-box without expensive training data collection. 
To enable such zero-shot cost models, we suggest a new learning paradigm based on pre-trained cost models. As core contributions to support the transfer of such a pre-trained cost model to unseen databases, we introduce a new model architecture and representation technique for encoding query workloads as input to those models.  As we will show in our evaluation, zero-shot cost estimation can provide more accurate cost estimates than state-of-the-art models for a wide range of (real-world) databases without requiring any query executions on unseen databases. Furthermore, we show that zero-shot cost models can be used in a few-shot mode that further improves their quality by retraining them just with a small number of additional training queries on the unseen database. 
\end{abstract}

\maketitle

%
%
%

\section{Introduction}
\label{sec:intro}

\begin{figure}
	\centering
	\includegraphics[width=0.9\columnwidth]{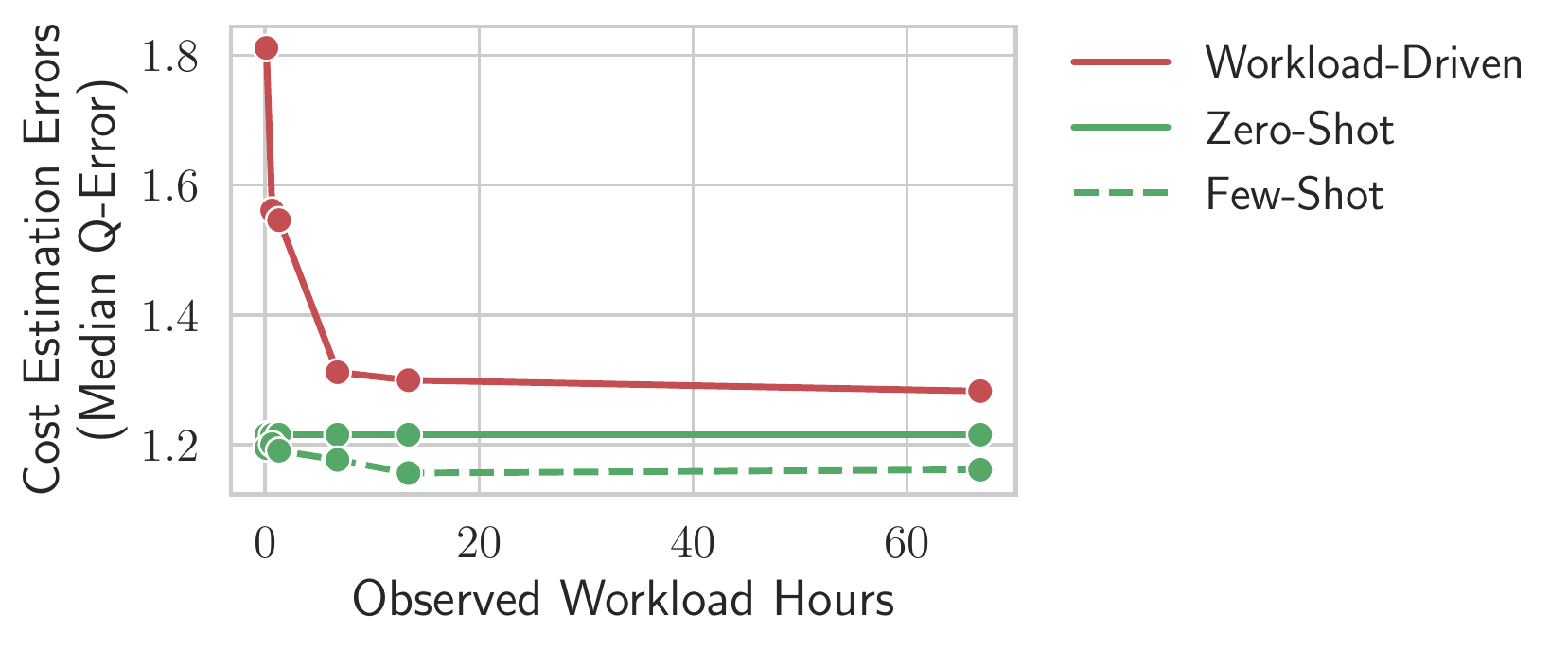}
	\vspace*{-2.5ex}
	\caption{Cost Estimation Errors on the IMDB database. While workload-driven approaches \cite{sun2019endtoend} require many hours of workload executions as training data, our zero-shot cost model supports the unseen IMDB database \textit{out-of-the-box} and provides highly accurate cost estimates. If a workload is observed however, these queries can be used to fine-tune a zero-shot model resulting in few-shot models which further improve the performance.}
	\vspace*{-3.5ex}
	\label{fig:motivation}
\end{figure}

\paragraph{Motivation.}
Accurate physical cost estimation (i.e., estimating query latencies) is crucial for query optimization in DBMSs.
Classically, cost estimation is performed using models that make several simplifying assumptions. As a result, such models often over- or underestimate runtimes, leading to suboptimal planning decisions that degrade the overall query performance \cite{leis2018looking}.
Recently, machine learning has thus been used for learned cost models that do not need to make such simplifying assumptions \cite{sun2019endtoend}. 

While it was shown that the cost estimates of such learned cost models are significantly more accurate than those of the traditional cost models, the existing approaches rely on \textit{workload-driven} learning where models have to observe thousands of queries on the same database\footnote{Throughout this paper, we use the term database to refer to a particular dataset with certain data characteristics.} for which the cost prediction should be performed. This workload execution is required to gather the training data which can take hours (or days) since tens of thousands of queries need to be executed on potentially large databases.

In Figure~\ref{fig:motivation}, we show the cost estimation accuracy depending on how many hours we allow for gathering the training data for a workload-driven model. As we can see, even for a medium-sized database such as IMDB, it takes more than $5$ hours of running queries on this database to gather enough training data such that the cost estimation model can provide a decent accuracy.

Unfortunately, collecting training data by running queries is not a one-time effort. In fact, the training data collection has to be repeated for every new database a learned model should be deployed for. This is due to the fact that current model architectures for workload-driven learning tie a trained model to a particular database instance.
Consequently, for every (new) unseen database we not only have to train a model from scratch but also gather training data in the form of queries. And even for the same database, in case of changed data characteristics due to updates, training data collection needs to be repeated. 
Overall, these repeated high costs for obtaining training data for unseen databases render workload-driven learning unattractive for many practical deployments.

\vspace*{-1.5ex}\paragraph{Contributions.}
In this paper, we thus suggest a new learning paradigm for cost estimation called zero-shot cost models that reduces these high efforts.
The general idea behind zero-shot cost models is motivated by recent advances in transfer learning of models.
Similar to other approaches such as GPT-3 \cite{NEURIPS2020_1457c0d6} which enable zero-shot learning for NLP, a zero-shot cost model is trained on a wide collection of different databases and workloads and can thus generalize \textit{out-of-the-box} to a completely unseen database without the need to be trained particularly on that database.
In fact as depicted in Figure~\ref{fig:motivation}, zero-shot cost models can provide a high accuracy and often even outperform existing workload-driven approaches that have been trained on large sets of training queries.
Moreover, as we also show in Figure~\ref{fig:motivation}, zero-shot cost models can additionally be fine-tuned on the unseen database with just a few training queries and the resulting few-shot models further improve the accuracy.

One could now argue that it might be a significant effort to collect sufficient training data across databases for pre-training a zero-shot model. However, in contrast to workload-driven models which require training data for every unseen database, training data collection is a one-time effort; i.e., once trained the zero shot model can be used for any new unseen database.
In fact, in our evaluation we show that zero-shot models can provide high accuracies for a wide-variety of real-world databases. 
Moreover, for historical traces can be used which eliminates the need to collect any training data.
For example, cloud providers such as AWS, Microsoft, or Google, typically anyway keep logs of their customer workloads which could directly be used as training data for zero-shot learning without collecting any further training data.

A key aspect to enable zero-shot learning is that a cost model can be transferred to new (unseen) databases, i.e., the models leverage observed query executions on a variety of different databases to predict runtimes on new (unseen) databases. However, state-of-the-art model architectures used for workload-driven learning do not support this training and inference mode since they are tied to a particular database.
As a core novel contribution for zero-shot cost models we thus devise a new model architecture based on a representation of queries that generalizes across databases using a \textit{transferable} representation with features such as the tuple width that can be derived from any database.
Moreover, zero-shot models \textit{separate concerns}; i.e., data characteristics of a new database (e.g., rows of tables) are not implicitly learned as in classical workload-driven learning (which hinders generalization), but are provided as input to the model.

Another core question for zero-shot models is at which point a sufficient amount of different training databases and workloads was observed to generalize robustly to unseen databases. To answer this question, as a second contribution in this paper we derive a method to estimate how accurate the runtime estimations of zero-shot models will be for unseen databases. We also discuss how to address cases of workload drifts where the zero-shot models are expected to generalize less robustly.
Furthermore, we also show that zero-shot models are widely applicable beyond cost models for query optimizers for single-node DBMSs which is the main focus of this paper. 
For instance, we believe that zero-shot cost models can be naturally extended to to distributed DBMS or even other use cases such as providing cost estimates for design advisors where the goal is to automatically find a suitable database design (e.g., a set of indexes) for a given workload. 

Finally, in our extensive experimental evaluation, we verify that zero-shot cost models generalize robustly to unseen databases and workloads while providing cost estimates which are more accurate than those of workload-driven models. As part of this evaluation, we also provide a new benchmark (beyond JOB) which is necessary to evaluate cost estimation models more broadly on a variety of (real-world) databases. We will make this benchmark including query executions for training cost models publicly available and hope that it will benefit future research in learned cost estimation and potentially beyond.

\vspace*{-1.5ex}\paragraph{Outline}
In Section~\ref{sec:overview}, we give an overview of our approach and describe the model architecture in more detail in Section~\ref{sec:zero-shot-models}. We then derive formal methods to recognize when sufficient training data is available for the model to generalize in Section~\ref{sec:generalization} and then discuss our extensions to show the broader applicability of zero-shot cost models in Section~\ref{sec:extensions}. Before discussing the evaluation in Section~\ref{sec:experiments}, we describe the design decisions for our proposed benchmark to evaluate cost models. Finally, we present related work (Section~\ref{sec:related-work}) and conclude in Section~\ref{sec:conclusion}. 

\section{Overview}
\label{sec:overview}

In this section, we introduce the problem of zero-shot cost estimation and then present an overview of our approach.

\subsection{Problem Statement}

The overall goal of zero-shot cost estimation is to predict query latencies (i.e., runtimes) on an unseen database without having observed any query on this unseen database. Throughout this paper we use the term database to refer to a particular dataset (i.e., a set of tables with a given data distribution). 
Note that the problem of zero-shot cost estimation is thus in sharp contrast to the problem addressed by state-of-the-art workload-driven cost models which train a model per database.
Finally, while we believe that zero-shot learning for DBMSs is more generally applicable, we restrict ourselves in this paper to cost estimations for relational DBMSs (both single-node and distributed). In particular, zero-shot cost models for other types of systems such as graph-databases or streaming systems are interesting avenues of future work.

\subsection{Our Approach}

A key challenge for developing zero-shot cost models is the question how to design a model that allows to generalize across databases. Here, we draw inspiration from the way classical cost models in DBMSs are designed. Typically, these consist of two models: a database-agnostic model to estimate the runtime cost and a database-dependent model (e.g., histograms) to capture data characteristics.
When predicting the cost of a query, the estimated cardinalities and other  characteristics (i.e., outputs of the database-dependent models) serve as input to the general database-agnostic cost model which captures the general system behavior (e.g., the costs of a sequential scan grows linearly w.r.t. the number of rows). While the classical models are lightweight, they often largely under- or overestimates the true costs of a query since models are too simple to capture complex interactions in the query plan and data.

\begin{figure}
	\centering
	\includegraphics[width=0.85\columnwidth]{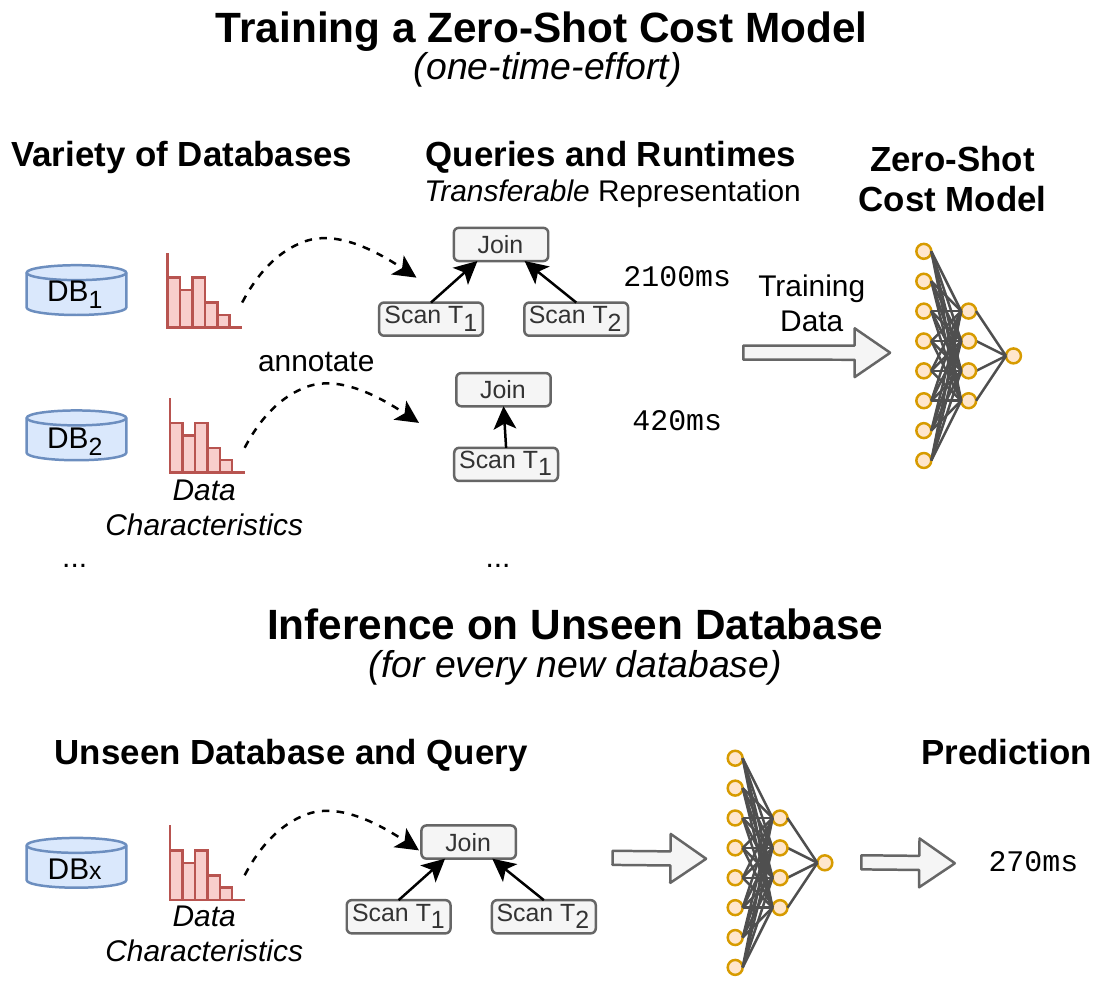}
	\vspace*{-2.5ex}
	\caption{Overview of Zero-Shot Cost Estimation. The zero-shot cost model is trained \textit{once} using a variety of queries and databases. At inference time, the model can then provide cost estimates for an unseen database and queries without requiring additional training queries. Enabling zero-shot cost estimation is based on two key ideas: (1) a new \textit{transferable} query representation and model architecture is used to enable cost predictions on unseen databases and (2) we separate concerns, i.e., a zero-shot model learns a general database-agnostic cost model which takes database-specific characteristics as input.
	}
	\vspace*{-3.5ex}
	\label{fig:overview}
\end{figure}

Hence, in our approach, we also separate concerns but use a much richer learned model which similarly takes data characteristics of the unseen database as input to predict query runtimes in a database-agnostic manner.
As depicted in Figure~\ref{fig:overview} (upper part), for training such a zero-shot cost model we provide different query plans along with the runtime as well as the data characteristics of the plan (such as tuple width as well as intermediate cardinalities) to the zero-shot cost model. Once trained, the model can be used on unseen databases to predict the query runtime as shown in Figure~\ref{fig:overview} (lower part).

As mentioned before, to predict the runtime of a query plan on a new (unseen) database, we feed the query plan together with its data characteristics into a zero-shot model.
While data characteristics such as the tuple width can be derived from the database catalogs, other characteristics such as intermediate cardinalities require more complex techniques.
To derive intermediate cardinalities of a query plan in our approach we thus make use of data-driven learning \cite{hilprecht2020deepdb,yang2020naru} that can provide exact estimates on a given database. Note that this does not contradict our main promise of zero-shot learning since data-driven models to capture data characteristics can be learned without queries as training data.

Another core challenge of enabling zero-shot cost models that can estimate the runtime of a plan given its data characteristics is how to represent query plans which serve as input to the model. While along with workload-driven cost models, particular representation methods for query plans have already been proposed, those are not applicable for zero-shot learning. 
The reason is that the representations are not \textit{transferable} across databases. 
For instance, literals in filter predicates are provided as input to the model (e.g., \texttt{2021} for the predicate \texttt{movie.production\_year=2021}). However, the selectivity of literals will vary largely per database since the data distribution will likely be different (e.g., there might not even exist movies produced in 2021 in the test database).

Hence, as a second technique in this paper, we propose a new representation for queries that completely relies on features that can be derived from any database to allow the model to generalize to unseen databases. 
For example, predicates for filter operations in a query are encoded by the general predicate structure (e.g., which data types and comparison operators are used in a predicate) instead of encoding the literals. In addition, data characteristics of a filter operator (e.g., input and output cardinality to express the selectivity) are provided as additional input to a zero-shot model.
That way, a zero-shot model can learn the runtime overhead of a filter operation based on database-agnostic characteristics.
We present details of our query representation in Section~\ref{sec:zero-shot-models}.

Finally, a last important aspect of zero-shot cost models is that they can easily be extended to \textit{few-shot learning}.
Hence, instead of using the zero-shot model out-of-the box (which already can provide good performance), one can fine-tune the model with only a few training queries on an unseen database.

 \section{Zero-Shot Cost Models}
\label{sec:zero-shot-models}

\begin{figure*}
	\centering
	\includegraphics[width=0.94\linewidth]{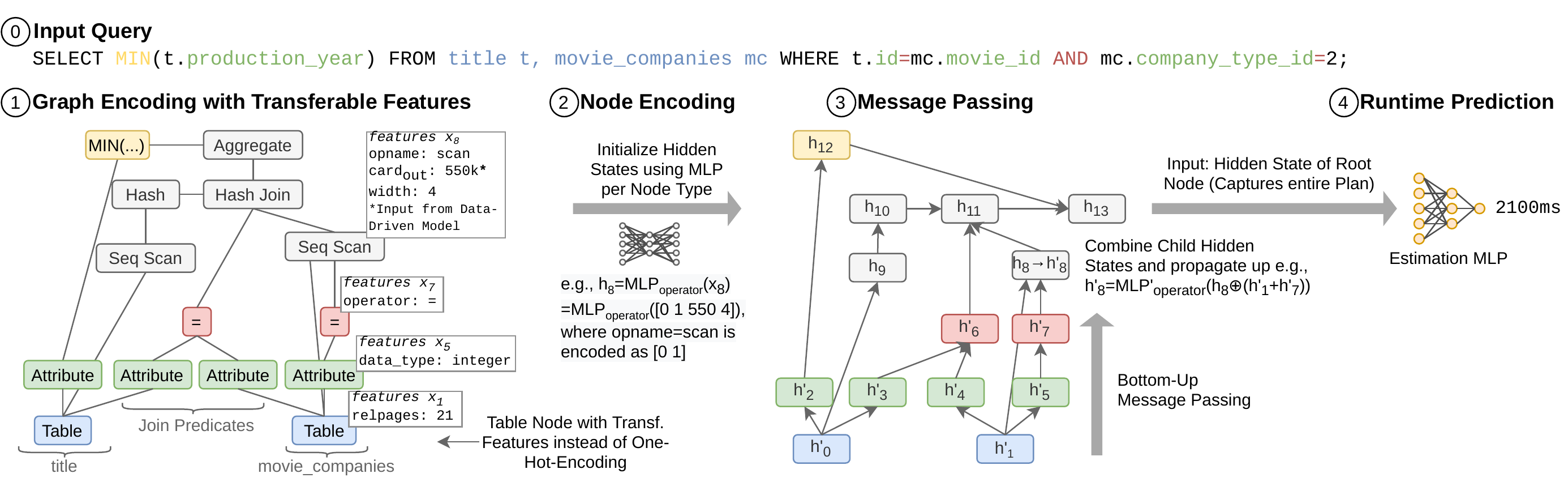}
	\vspace*{-3.5ex}
	\caption{Using Zero-Shot Models for Cost Estimation  (i.e., for Inference) on a unseen Database. (1) A query is represented as a graph with different node types (to represent plan operators, predicates, tables, attributes etc.) and nodes are annotated with \textit{transferable} features which  generalize across databases. (2) Afterwards, the resulting feature vectors of the nodes are fed into node-type-specific Multi-Layer-Perceptrons (MLPs) to obtain a hidden state which is (3) propagated through the query-tree using bottom-up message passing to account for interactions among connected nodes. (4) Finally, the hidden state of the root node (encoding the entire graph) is fed into a final model --- the estimation MLP --- which predicts the query runtime.}
	\label{fig:overview-zero-shot}
	\vspace*{-3.5ex}
\end{figure*}

As mentioned in Section~\ref{sec:overview}, a zero-shot cost model (once trained) is able to predict the runtime of a query on an entirely new database without retraining. 
A core building block needed to enable a zero-shot model is a new representation of queries that can generalize across databases.
In the following, we thus first explain how we devised such a transferable query representation and then discuss how inference and training of a zero-shot model that uses this representation works.

\subsection{Query Representation}

State-of-the-art workload-driven models \cite{kipf2019learned, sun2019endtoend} for cost estimation do not use transferable query representations and can thus only be used on the database they were trained on. To better understand why current query representations are not transferable, we first explain how they typically encode queries.

\vspace*{-1.5ex}\subsubsection{Query Representation for Workload-Driven Models.} At the core, query representations used for workload-driven approaches hard-code the model against a single database.
For example, attribute names (e.g., those used in filter predicates) are typically encoded using a one-hot encoding assigning each attribute present in the database a specific position in a feature vector. For instance, the attribute \texttt{production\_year} of the IMDB dataset might be encoded using the vector $(0,1,0)$ (assuming that there are only three attributes in total). If the same model should now be used to predict query costs for the SSB dataset, some attributes might not even exist or even worse they might exist but have very different data distributions or even a different data type. In fact, non-transferable feature encodings are not only used for attributes but in various places of the query representation such as encoding table names or literals in filter predicates.

\vspace*{-1.5ex}\subsubsection{Query Representation for Zero-Shot Cost Models.}
Hence, for zero-shot cost models we require a new query representation that is transferable across databases. The main idea of the transferable representation we suggest in this paper is shown in  Figure~\ref{fig:overview-zero-shot}. 
At the core, a query plan and the involved tables and attributes are represented using a graph where graph nodes use transferable features \circled{1} (i.e., features that provide meaningful information to predict runtime on different databases).
This representation then serves as input for the training and inference process of zero shot cost models \circled{2}-\circled{4} that we explain in the subsequent sections.
In the following, we discuss the graph encoding of the transferable featurization in detail.

\vspace*{-1.5ex}\paragraph{Graph Encoding of Query Plans.} 
While graph-based representations have been already used to represent query operators of a query plan \cite{sun2019endtoend}, our representation has significant differences.
First, as shown in Figure~\ref{fig:overview-zero-shot} \circled{1}, our representation not only encodes physical plan operators as nodes (gray) in the graph as in previous work \cite{sun2019endtoend}, but it also covers all query plan information more holistically using different nodes types for attributes (green), tables (blue) as well as predicate information (red).
Second, as discussed before, previous approaches also covered such information, however, they used one-hot-encodings (which are non-transferable) while our representation captures the query complexity in a transferable way. 

For instance, to encode filter predicates, different from previous approaches we encode the predicate structure as nodes (red) without literals. In particular, we encode information such as data types of the attributes and operators used for comparisons.
For example, the filter predicate \texttt{company\_type\_id=2} for the query \circled{0} in Figure~\ref{fig:overview-zero-shot}, is encoded using an \emph{attribute} node ($x_5$) with the comparison node \emph{=} ($x_7$).
As such, a zero-shot cost model provided with the transferable features (e.g., intermediate cardinalities which are given by the data-driven models) can infer the complexity of the predicates to estimate the query runtime.

\vspace*{-1.5ex}\paragraph{Transferable Featurization.} 
While our graph representation allows to flexibly encode query plans across databases, we similarly have to make sure that the features used to represent nodes in the graph \circled{1} (e.g., plan operators as shown in gray) are transferable. 
In particular, when used on different databases, features should not encode any implicit information that hinder the transfer of the model to a new unseen database.
The set of such features used for the different node types in our query representation is depicted in Table~\ref{tab:features}. 
For instance, attribute nodes (green) use features such as the data type or the width in bytes.
Similarly, for tables (blue nodes), we use other transferable features (e.g., the number of rows as well as the number of pages on disk). 

Importantly, transferable features can either characterize the query plan (e.g., operator types) or represent the data characteristics (e.g., intermediate cardinalities) and together allow a zero-shot cost model to generalize to an unseen database. 
For transferable features that represent data characteristics many can be derived from the metadata of a database (such as the the number of rows of a table node).
However, some other features that represent data characteristics --- e.g., the estimated output cardinality of an operator node --- require more involved techniques.  
In Section~\ref{sec:data-characteristics}, we discuss alternatives of how we provide estimated output cardinalities to zero-shot cost models.

\begin{table}
	\scriptsize
	\centering
	\begin{tabularx}{8cm}{llX}
		\toprule
		Node Type & Feature & Description\\ 
		\midrule
		Plan Operator & \texttt{card}$_\texttt{out}$ & Estimated output cardinality of operator \\
		& \texttt{width} & Tuple width \\
		& \texttt{workers} & Number of parallel workers \\
		& \texttt{opname} & Name of physical operator \\
		& \texttt{card\_prod} & Estimated product of children output cardinalities \\
		Predicate & \texttt{operator} & Operator type (e.g., \texttt{=}) \\
		& \texttt{literal\_feat} & Feature capturing literal complexity, e.g., number of values for \texttt{IN} operator or regex complexity \\
		Table & \texttt{relpages} & Number of pages \\
		& \texttt{reltuples} & Number of rows\\
		Attribute & \texttt{width} & Avg. number of bytes to represent a value \\
		& \texttt{correlation} & Attribute correlation with row number \\
		& \texttt{data\_type} & Data type of column \\
		& \texttt{ndistinct} & Number of distinct attribute values\\
		& \texttt{null\_frac} & Fraction of \texttt{NULL} values\\
		Output Column & \texttt{aggregation} & Which aggregation type is used\\
		\bottomrule
	\end{tabularx}
	\caption{Zero-Shot Features. All features are transferable and have the same semantics for different databases.}
	\vspace*{-7.5ex}
	\label{tab:features}
\end{table}

\subsection{Inference on Zero-Shot Models} 

Once a query graph with the transferable features on a unseen database is constructed for a query plan, it can be used as input for a (trained) zero-shot cost model to predict the runtime. 
Predicting the runtime of a new query plan with a zero-shot cost model is executed in three steps which we depict as pseudocode in Algorithm \ref{alg:forward_prop}: First, we compute a hidden state for every node of the query graph \circled{2} given the node-wise input features. Second, the information of different graph nodes is combined using message passing \circled{3} before a Multi-Layer-Perceptron (MLP) predicts the runtime of the query plan \circled{4}.
The same steps are also reflected in Figure~\ref{fig:overview-zero-shot} \circled{2} to \circled{4}.

In particular, in step \circled{2}, the feature vectors $x_v$ of each graph node $v$ are encoded using a node-type specific MLP, i.e., nodes of the same type (e.g., all plan operators) use the same MLP to initialize their hidden state $h_v$ (line~\ref{alg:line:hiddenstate}).  For instance, in Figure~\ref{fig:overview-zero-shot}, the hidden state $h_{8}$ of the node representing the sequential scan on the \texttt{movie\_companies} table is obtained by feeding the feature vector $x_{8}$ (containing transferable features) into an MLP which is shared among all plan operators (gray nodes).

Afterwards, in step \circled{3}, a message passing scheme is applied which is prominently used in graph neural networks (GNNs) \cite{pmlr-v70-gilmer17a} to model the interactions between nodes in graphs (i.e., to capture interactions of query operators in the plan such as effects of a pipelined query execution). 
Different from message passing schemes for general graph encodings, for the message passing in zero-shot models we can exploit the fact that queries can be represented as directed acyclic graphs (DAGs) since query-plans are tree-structured. We thus use a novel bottom-up message passing scheme through the graph (i.e., in topological ordering) to obtain an updated hidden state $h'_v$ of a node $v$ that contains all information of the child nodes. During this pass, the updated hidden states $h'_u$ of the children $u$ are combined by summation \cite{NIPS2017_f22e4747} and concatenated with the initial hidden state $h_v$ of a node and fed into a node-type-specific MLP (line~\ref{alg:line:combine}). 
For instance, in Figure~\ref{fig:overview-zero-shot}, the updated hidden state $h'_{8}$ of the scan node is obtained by summing up the updated hidden states of the child nodes (representing the table and predicate operator of the scan) concatenated with the initial hidden state (capturing properties of the scan) which is then fed into an MLP which is again shared among all plan operators.

Finally, as a result of step \circled{3} the updated hidden state $h'_r$ of the root node $r$ of a query plan captures the properties of the entire query. For the cost prediction in step \circled{4}, we thus feed this hidden state into a final estimation MLP to obtain the cost estimate $\hat{c}=MLP_{\mathit{est}}(h_r)$ (line~\ref{alg:line:estimate}). Hence, in Figure~\ref{fig:overview-zero-shot} \circled{4}, the updated hidden state $h'_{13}$ is fed into the final estimation MLP to obtain the cost estimate since it captures information of the entire plan.

\begin{algorithm}[t]
	\scriptsize
	\caption{Inference on Zero-Shot Models}
	\label{alg:forward_prop}
	\begin{algorithmic}[1]
		\State \textbf{Input:} Query graph encoding with nodes $v$ and input features $x_v$ 
		\State \textbf{Output: } Cost estimate $\hat{c}$ 
		\State		
		\For{$v\in$ graph encoding} \Comment{Compute hidden state per node \circled{2}}
		\State $h_v \leftarrow \mathit{MLP}_{T(v)}(x_v)$  \label{alg:line:hiddenstate}
		\EndFor
		
		\For{$v\in$ in topological ordering} \Comment{Bottom-up pass in graph \circled{3}}
		\State $h'_v$ $\leftarrow$ $MLP'_{T(v)}\left(\sum_{u\in\mathit{children(v)}}h'_u\oplus h_v\right)$  \label{line:child_agg} \label{alg:line:combine}
		\EndFor
		\State $\hat{c}\leftarrow MLP_{\mathit{est}}(h'_r)$ \Comment{Estimate costs using root node state \circled{4}} \label{alg:line:estimate}
		\State \Return $\hat{c}$
	\end{algorithmic}
\end{algorithm}

\subsection{Training Zero-Shot Models} 

As mentioned before, a zero-shot cost model is trained on several databases and queries to learn the runtime complexity of query plans given the input features.
To be more precise, a zero-shot cost model is trained in a supervised fashion using pairs $(P,c)$ that consists of a plan $P$ with the respective features and the actual runtime cost $c$. 
Importantly, all steps described in the inference procedure (node encoding, message passing and finally runtime estimation) are differentiable, which allows us to train the model parameters of the MLPs used for all zero-shot model components jointly in an end-to-end fashion. As loss function to compare the actual costs $c$ of a featurized query plan $P$ with the estimated costs $\hat{c}$, we use the Q-error loss $\max(\frac{c}{\hat{c}},\frac{\hat{c}}{c})$ \cite{kipf2019learned, sun2019endtoend} since this worked best for zero-shot models compared to other alternatives.

\subsection{Deriving Data Characteristics}
\label{sec:data-characteristics}

As discussed before, an important aspect of a zero-shot model is that the model is not tied to a particular data distribution of a single database.
For enabling this, we provide data characteristics such as attribute widths in bytes, number of pages and tuples of tables but also output cardinalities of operators as input to those models.
To be more precise, given a particular query plan for which the runtime should be estimated, those features have to be annotated for each graph node in the query encoding. 

While the majority of those features can simply be derived from the database catalog, intermediate cardinalities in a query plan are notoriously hard to predict and simple statistics are known to be often imprecise \cite{leis2018looking}.
Hence, learned approaches to tackle cardinality estimation have been proposed to derive accurate intermediate cardinalities. 
While in principle such learned approaches can be used to predict intermediate cardinalities which are then used as input for the zero-shot models, there are important trade-offs when choosing which techniques are suitable for zero-shot learning (cf. Table~\ref{tab:cardinality-approaches}). In the following, we discuss these aspects.

\begin{table}
	\scriptsize
	\centering
	\begin{tabularx}{8.5cm}{lXXXX}
		\toprule
		& Zero-Shot Compatible & Accurate Estimates & Training Overhead  & Arbitrary Queries \\ 
		\midrule
		Traditional & \OK & & \emph{No} & \OK \\
		Workload-Driven  & & \OK & \emph{High} &  \\
		Data-Driven  & \OK & \OK & \emph{Low} &  \\
		\bottomrule
	\end{tabularx}
	\caption{Trade-offs of different Cardinality estimators used with Zero-shot Models. Data-driven models are a promising choice offering low overhead and high accuracy.}
	\vspace{-7.5ex}
	\label{tab:cardinality-approaches}
\end{table}

First, a zero-shot cost model should be able to predict query runtimes on databases that were not seen before without relying on an observed workload on that database. Since workload-driven models for cardinality estimation require such queries as training data, they are not suited for our purpose of predicting cardinalities for zero-shot models. Second, traditional histogram-based approaches have the advantage that no additional efforts are required since the query optimizers anyway have built-in techniques. However, they are often imprecise. Third, data-driven models are more precise but also need to be trained. However, the training does not rely on query executions and is thus usually just in the order of minutes. Unfortunately, state-of-the-art data-driven cardinality estimators do not yet support the same variety of different queries as traditional approaches. 

Depending on whether the effort to train a data-driven model for an unseen database is acceptable and the workload is supported, one can either opt for traditional approaches or data-driven models. Hence, we propose to use zero-shot models with data-driven cardinality models if possible and use optimizer cardinality estimates only as a fallback.
In our evaluation, we see that zero-shot models can still produce reasonable estimates even if only cardinalities estimates from traditional models are available.

\section{Robustness of Zero-Shot Models}
\label{sec:generalization}

An important question for zero-shot models is at which point a sufficient amount of different training databases (and workloads) was observed to generalize robustly to unseen databases. To answer this question, we first derive a method to estimate how robust the runtime estimates of zero-shot models will be for unseen databases. We then discuss a simple method to detect cases of workload drifts (i.e., the queries at runtime have substantially different characteristics than the training queries) as well as strategies how to tackle this problem.

\subsection{Estimating the Generalization Performance}
\label{sec:estimating_generalization}

We first formalize the problem, before we derive a method to estimate the generalization error. For training a zero-shot model, we have observed $n$ databases and workloads. In particular, for each of the databases $D_i$ we have access to training data $T_i$ in the form of query plans and their runtimes $T_i=\{(P_1,c_1),(P_2,c_2),\dots(P_m,c_m)\}$. We are now interested in how accurately the zero-shot cost model $Z$ will predict the runtimes for plans $T^*$ on some unseen database $D^*$. In particular, if the expected error is acceptable, we have observed a sufficient amount of databases and workloads.
More formally, we will define some error metric $E(T_i)$ with which we can compare the true runtimes and model predictions for some database $D_i$. An example for such a metric could be the prominently used median Q-error. We are now interested in estimating this error metric for an unseen database $E(T^*),$ i.e., the expected generalization error.

We now make use of statistical techniques to estimate the generalization error. For instance, in ML it is standard practice to train the model on a subset of the data and then use the remaining samples to estimate the error for future unseen datasets. Analogously, we can train the zero-shot model on a subset of the training databases $T_1,T_2,\dots,T_i$ (i.e., for a subset of databases) and evaluate the trained model on the remaining databases $T_{i+1},\dots,T_n$. Similar to cross validation, we can repeat this procedure with different splits and average the test errors to estimate the generalization error $E(T^*)$, i.e., how accurate the model is expected to be on an unseen database. Interestingly, this is an unbiased estimator of the test error $E(T^*)$ under the independent identically distributed (i.i.d.) assumption which we will discuss shortly. Hence, using only the observed databases and queries, we can estimate how accurate the model predictions for unseen databases will be. 

In order to now evaluate whether the model has observed a sufficient amount of databases and workloads, we can use two techniques. First, we can simply estimate the generalization error as described above and stop the training if it is sufficient. However, in this case we have to decide which generalization error is acceptable. A second technique (which we actually use in this paper) is to estimate if additional training databases will improve the generalization performance. For this, we train the model on subsets of all training databases. If the estimated generalization error $E(T^*)$ does not improve significantly for a larger number of training databases, we can conclude that additional databases will not improve the generalization capabilities of the zero-shot cost model and thus stop the training data collection.

\subsection{Tackling Workload Drifts}

The performance of zero-shot cost models will deteriorate if the new database and workload is significantly different from the training data. For instance, if there are significantly larger joins in the unseen database than for the training databases, the zero-shot model might not be able to predict the runtime with the same high accuracy. 
As we will show in our experimental evaluation, however, zero-shot cost models can often still generalize robustly in practice and can provide more accurate estimates than other baselines in case of workload drifts.
In addition, we suggest a strategy to detect cases of workload drifts by monitoring the test error and propose to tackle workload-drifts using few-shot learning.

Note that in cases of workload drifts the i.i.d. assumption does not hold and the Q-error on the unseen database is larger than implied by the generalization error. More technically, the i.i.d. assumption is a common assumption in ML that requires that the training datasets and test datasets are independent samples of some distribution $\mathcal{D}$. Due to a workload drift, the samples are no longer independent and thus the generalization error $E(T^*)$ might be increased. 
A simple yet effective strategy to recognize those cases is thus to monitor the error for unseen databases during inference. In cases where the error exceeds a certain threshold, one could decide to fine-tune the zero-shot model using the additional observed queries as training data (resulting in few-shot models). We will demonstrate in the experimental evaluation that zero-shot cost models fine-tuned on a small number of additional queries can significantly improve the accuracy on the unseen database in such cases.

\section{Extensions of Zero-Shot Cost Models}
\label{sec:extensions}

We now describe how zero-shot cost estimation can be extended in various directions.

\subsection{Distributed DBMSs}
\label{sec:distributed_dbms}

While the zero-shot cost models we discussed so far are centered around single-node DBMSs, we argue that zero-shot models can also be adapted to other types of DBMS. 
An important class of DBMSs are distributed DBMSs that are often found in the cloud to support scalable data processing.
To show the general feasibility, we now discuss some concrete extension of our approach to support a (commercial) cloud DBMS for OLAP .
Overall, we think that this demonstrates the potential of our approach to be adapted to new domains with the zero-shot paradigm in mind.

For supporting zero-shot cost estimates on the concrete cloud DBMS we used, two extensions were needed:
First, as mentioned before, cloud DBMSs are typically distributed and thus frequently shuffle the data during query execution (e.g., for distributed join processing). Second, independent of the fact that the processing is distributed there are often other optimizations to reduce the query execution costs of reading large amounts of data such as using a column store layout instead of a row store layout. 

\begin{figure}
	\centering
	\includegraphics[width=0.7\columnwidth]{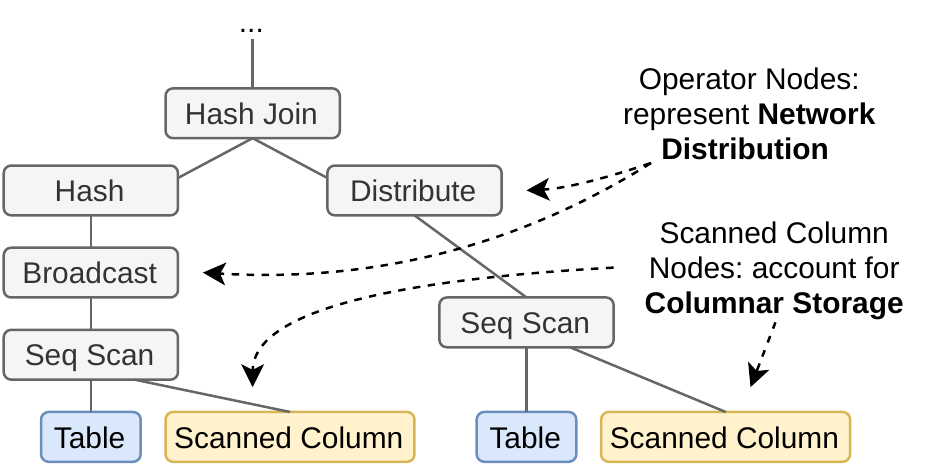}
	\vspace*{-2.5ex}
	\caption{Extension of Zero-shot Models to Distributed Cloud Data Warehouses. The columnar storage is accounted for by adding only nodes representing scanned columns to the graph whereas network shuffle operations (such as broadcast, partition by key etc.) required for distributed join execution are represented as physical plan nodes.}
	\label{fig:extensions}
	\vspace*{-3.5ex}
\end{figure}

To support these two aspects in zero-shot cost models, we extended the encoding of queries which serves as input to the zero-shot model as depicted in Figure~\ref{fig:extensions}. 
In particular, we included operator nodes for data shuffling as well as encode data formats (column or row) as a feature of the table node.

\subsection{Physical Design Tuning}

Another interesting direction is to use these models to estimate the runtime of query workloads for different (potential) physical designs.
This is helpful to automate physical design tuning which needs to enumerate different alternative designs and estimate which design to pick based on cost estimates (representing the runtime of a workload on a unseen physical design).

To show that zero-shot cost models support this, we extended the zero-cost models to be able to predict the runtime for queries with and without certain indexes on an unseen database.
For this, the query representation was extended such that the model can learn the trade-offs between different operations (e.g., how expensive a sequential scan is vs. an index scan).
Moreover, when training a zero-shot cost model, the training databases should include tables with and without indexes such that the model can learn how the cost for these two cases differ based on data characteristics.

In our experimental evaluation, we show that the zero-shot cost models can thus generalize robustly to unseen physical designs (i.e., to estimate the cost of a query with and without an index). However, there is clearly more to be done. 
For example, it could be beneficial to also introduce additional (transferable) features to capture other aspects of a physical designs (such as the expected height of indexes) or even support other options of physical designs (e.g., materialized views) in zero-cost models.

\vspace{-1.5ex}\subsection{Other Directions}

We believe that many more aspects such as hardware parameters (e.g., amount of available memory) and database knob configurations could be captured using zero-shot models.
Since those features are naturally transferable, e.g., knobs such as buffer sizes carry the same semantics across databases, an extension of our models will be possible. 
That way zero-shot models could inform also automated knob tuning.
However, while all these directions are interesting, they are beyond the scope of this paper and represent avenues of future work.

\section{A New Benchmark}
\label{sec:benchmark}

In order to properly train and evaluate cost models, we require both a diverse set of databases and executed workloads on these databases. Since currently there is no suitable benchmark with such properties, we created a new benchmark (that includes existing benchmarks such as JOB) which we discuss in this section. Furthermore, we will make this benchmark publicly available to foster future research in this area.

\subsection{Design Decisions}

For many years, DBMS systems were evaluated using synthetic benchmarks such as TPC-H \cite{tpch}, TPC-DS \cite{nambiar2006tpcds} or SSB \cite{o2009star}. While such benchmarks allow to evaluate the general system performance and scalability, they are in isolation insufficient to evaluate cost prediction models since the predicted cardinalities of the query optimizer are significantly more accurate than in practice. The reason is that the data is synthetic and thus no interesting correlations have to be captured making cardinality estimation challenging in practice. Hence, \citet{leis2018looking} suggested the JOB-workload on the IMDB dataset that comes with challenging correlations and has become the standard method (along with the simplified JOB-light workload \cite{kipf2019learned}) to evaluate learned cost and cardinality estimation approaches. 

While the IMDB benchmark is useful to evaluate workload-driven cost estimators that need to work on a single database only, it cannot be used for the evaluation of zero-shot cost models since these have to be trained on a variety of different databases. 
Moreover, even for workload-driven cost estimators a benchmark that spans a more diverse set of databases would definitively be helpful to evaluate the prediction quality.
Hence, we decided to create a new benchmark that covers established datasets such as IMDB but also additional datasets that have other characteristics.

\vspace{-1.5ex}\subsection{Datasets}

As discussed before, it is insufficient to just add synthetic datasets since correlations hardly resemble data distributions found in the real-world.
We thus decided to leverage  \emph{publicly available real-world datasets} \cite{motl2015ctu} together with the \emph{datasets used in established benchmarks such as JOB}. Since certain database were very small in size, we additionally scaled them to larger sizes to be interesting for cost estimation (s.t. a sample of queries takes a predefined threshold of time). In addition to the datasets mentioned before, we also included standard  benchmarks such as SSB and TPC-H to the benchmark.
As these benchmarks rely on synthetic data, this further increases the variety of data characteristics our benchmark covers for testing learned cardinality estimators. Overall, the benchmark comprises of 20 databases that vary largely in the number of tables, columns and foreign-key relationships.

\vspace{-1.5ex}\subsection{Workloads and Traces}

Furthermore, for benchmarking learned cost models, workloads are required for training and testing.
To simplify the comparison with prior work we first include predefined benchmark queries for databases that come with such workloads (e.g., JOB for IMDB).
However, since for the majority of the databases mentioned before no workloads are available, we implemented a workload generator that generates different types of queries.
For creating the workload, the generator supports three modes: a \textit{standard} mode where Select-Project-Aggregate-Join (SPAJ) queries with conjunctive predicates on numeric and categorical columns similar to the ones used by \citet{kipf2019learned} are generated, a more \textit{complex} mode which includes predicates involving disjunctions, string comparisons with regex predicates, \texttt{IS (NOT) NULL} comparisons and \texttt{IN} operators (resembling the complexity of the JOB-workload) and finally an \textit{index} workload where random indexes (both foreign key and for predicate columns) are created throughout the execution of the standard workload which is challenging due to the varying physical designs. Since the benchmark will be publicly available it can be easily extended to support an even broader class of queries in the future.

In addition to the datasets and the workload generator, the benchmark comes with workload traces (e.g., executions of the queries and their runtime) for all 20 databases that can be used directly by other researchers as training / testing data (which we also used in our evaluation).
To be more precise, we generated $15,000$ queries per database and executed those queries on a Postgres DBMS (v12) on \texttt{c8220} nodes on the cloudlab platform. 
Overall, this also allows for a better reproducibility since this platform can be used by other researchers as well.
To limit the already excessive resource consumption required to produce this trace, we excluded queries running longer than 30 seconds from the benchmark. In total, the execution of these more than 300k queries takes 10 days if executed on a single node. As part of the traces, we not only provide the runtime of the query but also the physical plan used to run the query along with actual cardinalities.

\section{Experimental Evaluation}
\label{sec:experiments}

In this Section, we evaluate zero-shot cost estimation with a set of different experiments:

\begin{itemize}
	\item \textbf{Exp 1. Zero-Shot Accuracy.} We evaluate how accurately zero-shot cost models can predict costs for unseen databases.
	\item \textbf{Exp 2. Zero-Shot vs. Workload-Driven.} In addition, we compare the training overhead and accuracy with state-of-the-art workload driven approaches.
	\item \textbf{Exp 3. Generalization.} In this experiment, we study how our models generalize under workload drifts (i.e., under database updates and larger unseen joins).
	\item \textbf{Exp 4. Extensions.} We then study the broad applicability of zero-shot cost models beyond single-node cost estimation (i.e., for distributed DBMSs and different physical designs).
	\item \textbf{Exp 5. Training and Inference Performance.} Furthermore, we evaluate the training and inference performance of zero-shot cost models and compare training efforts to workload-driven models.
	\item \textbf{Exp 6. Ablation Study.} Finally, we show the effects of different design alternatives of zero-shot models as well as a study where we determine how many database are sufficient for zero-shot cost models to generalize.
\end{itemize}
\vspace*{-1ex}
	
For all experiments, we use the traces of the benchmark discussed before (for training and testing).

\subsection{Exp 1: Zero-Shot Accuracy}
\label{sec:exp_zero_shot_eval}

\begin{figure}
	\centering
	\includegraphics[width=0.95\linewidth]{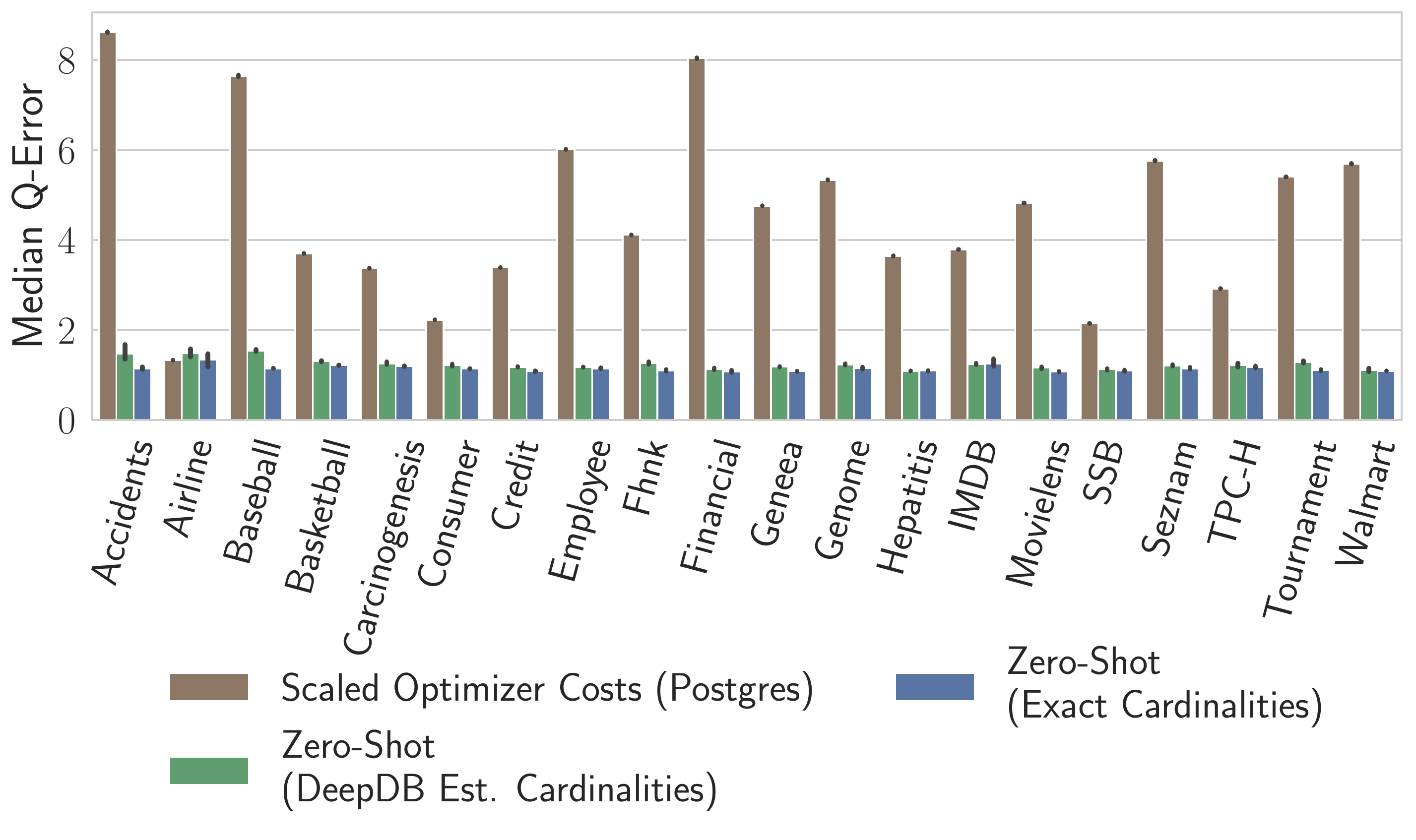}
	\vspace*{-3.5ex}
	\caption{Zero-Shot Generalization across Databases. The zero-shot models are trained using workloads on 19/20 databases and tested on the remaining unseen database. Overall, zero-shot models are significantly more accurate than the scaled estimates of the optimizer cost model. 
}
	\label{fig:zero_shot_gen}
	\vspace*{-4.5ex}
\end{figure}

First, in order to evaluate the accuracy of zero-shot cost models, we trained a zero-shot model using workloads on $19$ out of the $20$ datasets of the benchmark as training data and evaluated the model on the workload of the unseen (remaining) database. In particular, we use the trained model to predict the runtimes of the queries on the unseen database and report the median Q-error. 
In the first experiment, we focus on the \textit{standard} workloads and defer the results of the \textit{complex} and \textit{index} workloads of our benchmark to follow-up experiments. For this experiment, we repeat the cost estimation for every unseen database with three runs using different seeds 

For showing the performance of zero-shot cost models on unseen databases, we used two variants of providing intermediate cardinalities - we either used predictions of learned cardinality estimators or the actual cardinalities which are not available in practice prior to execution but serve as an interesting upper baseline for zero-shot learning (i.e., how accurate the predictions become with perfect cardinality estimates). For the data-driven cardinality estimator, we trained DeepDB \cite{hilprecht2020deepdb} models, which worked best in preliminary experiments.

To the best of our knowledge, we are the first to propose zero-shot cost estimation and thus no other learned approaches are included as a direct baseline in this first experiment where we aim to analyze the accuracy on unseen databases. For instance, workload-driven approaches would need query executions on the unseen database which we do not provide in the zero-shot setting. However, we compare our approach with workload-driven models in Section~\ref{sec:exp_comp_wl_driven}. 

As a sanity check that zero-shot models provide better performance than classical cost estimation models that rely on simple (non-learned) techniques (and as such could also count as zero-shot cost models), we use cost estimates coming from the Postgres query optimizer as a baseline similar to previous work \cite{sun2019endtoend}. Moreover, for the distributed setup we later on also employ the cost estimates of a commercial cloud DBMS. 
Since Postgres cost estimates are provided as abstract cost units, we use a simple linear model on top of Postgres estimates (and hence the results are called \emph{Scaled Optimizer}) which provides actual query runtimes. 
Different from \cite{sun2019endtoend} which directly take the cost units as runtime (in ms), using a linear model on top results in a much lower Q-error for Postgres. 
For training the simple linear model we are using the same training data from the other $19$ databases as for zero-shot models to be fair.

The results can be seen in Figure~\ref{fig:zero_shot_gen}. In general, the zero-shot models offer robust performances for all of the databases despite the varying complexity. In fact, all median Q-errors are below $1.54$ for the version using DeepDB cardinality estimates (vs. $8.62$ in the worst case for the \emph{Scaled Optimizer} cost). Finally, we can see that zero-shot cost models using DeepDB cardinalities are almost matching the performance with perfect cardinalities. This suggests that the models can cope with partially inaccurate cardinalities. Indeed, as we will see in a follow-up experiment, this even holds when we use potentially inaccurate cardinality estimates coming from a classical the optimizer instead.

Overall, we can see that the zero-shot cost models are significantly more accurate than the scaled optimizer estimates outperforming these on $18$ out of $19$ datasets and being on par for the last remaining dataset (Airline). The reason is that zero-shot cost models capture subtleties in operator performance and interactions of operators in the plan more accurately than simplistic cost models. The results are just on par for the remaining database since the optimizer costs are relatively accurate because it is merely a star schema, i.e., a relatively simple schema structure.

\begin{figure}
	\centering
	\includegraphics[width=0.99\linewidth]{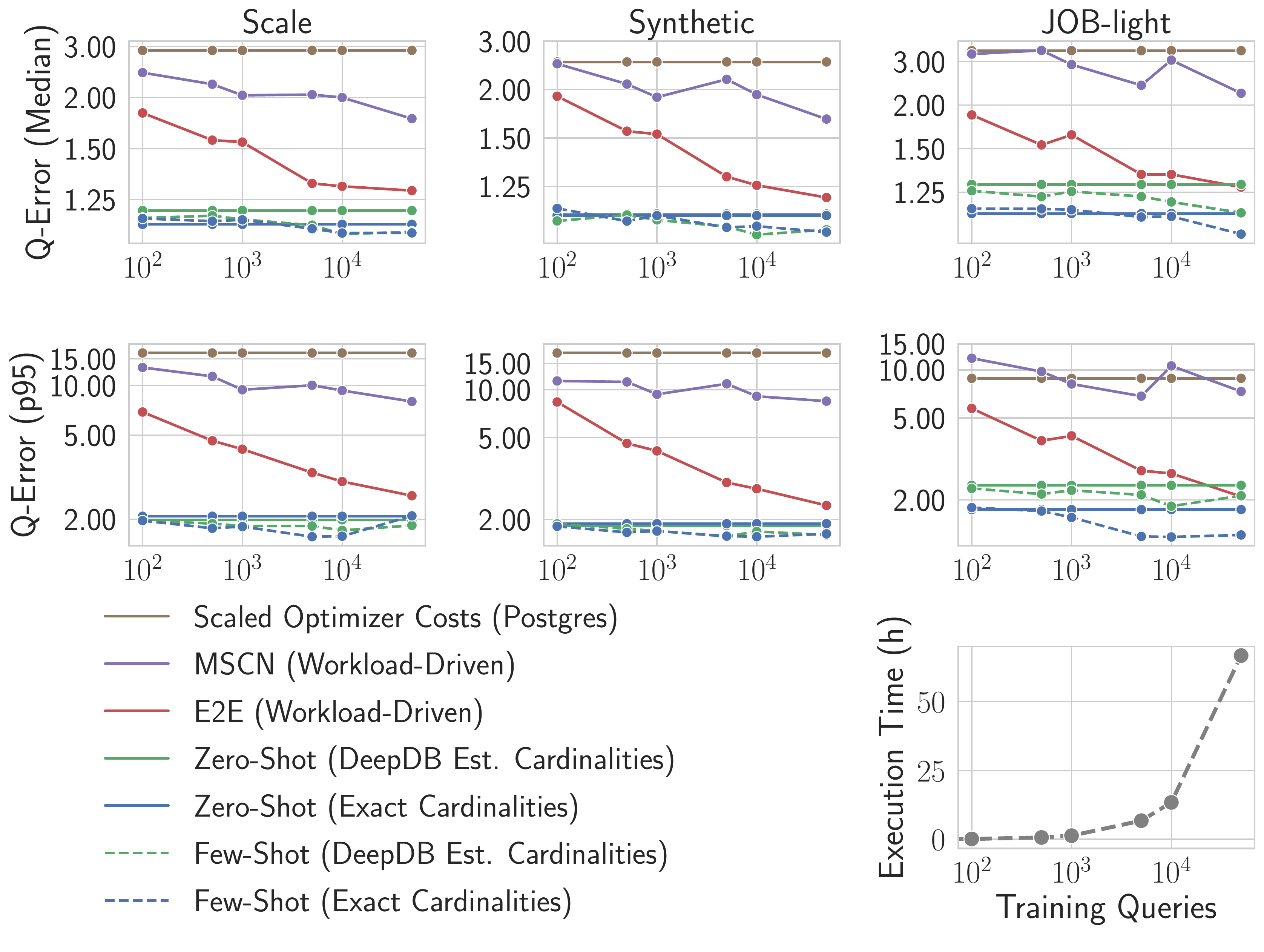}
	\vspace*{-2.5ex}
	\caption{Estimation Errors of Workload-Driven Models for a varying Number of Training Queries compared with Zero-Shot Cost Models. Even the most accurate workload-driven model (E2E) requires approximately 50k query executions on an unseen database for a comparable performance with zero-shot models which is roughly equivalent to 66 hours of executed workload. Since zero-shot models do not require any additional queries it is significantly cheaper to deploy them for a new database. However, zero-shot models can be fine-tuned to obtain few-shot models which further improve the accuracy.}
	\vspace*{-3.5ex}
	\label{fig:wl_driven_vs_zero_shot}
\end{figure}

\subsection{Exp 2: Zero-Shot vs. Workload-Driven}
\label{sec:exp_comp_wl_driven}

In the following, we contrast the performance of zero-shot cost models with workload-driven approaches.

\vspace*{-1.5ex}\paragraph{Training Overhead.} An interesting question is  how many training queries are required for workload-driven learning on an unseen database to match the performance of zero-shot learning which we will study next.
In particular, in this experiment we evaluate the Q-errors for the standard workloads (scale, synthetic, and JOB-light) on the IMDB database. As before zero-shot models are not trained on IMDB at all (but on the other $19$ databases) while workload-driven models are trained on a varying number of training queries on IMDB. 

For the workload-driven approaches we use the E2E model proposed by \citet{sun2019endtoend} as well as the MSCN model by \citet{kipf2019learned}. The idea of the E2E models is to featurize the physical query plans and feed them into a neural model to predict the runtime. However, in contrast to zero-shot cost models the query plan representation is not transferable and thus the train and test databases have to be identical. The MSCN model which was initially developed for cardinality estimation uses a more high level representation and encodes the sets of joins, predicates and tables of a query which are then fed into a neural architecture which is thus oblivious of the physical plans used. Both models are trained on a varying number of training queries which are generated for the IMDB dataset similar to the original training setup used by \citet{sun2019endtoend}. 
Furthermore, as a last baseline, we again employ the scaled costs of the Postgres query optimizer. 

In Figure~\ref{fig:wl_driven_vs_zero_shot}, we depict the median Q-error of comparing our zero-shot performance to the baselines as discussed before for the IMDB benchmark workloads for a varying number of training queries. 
As we can see the zero-shot cost models can estimate the runtimes accurately even though queries on the IMDB dataset were not observed in the training data. In particular, E2E requires about 50k training queries on the IMDB database to be on-par with zero-shot cost models. As we can see in the lower right plot in Figure~\ref{fig:wl_driven_vs_zero_shot} this amount of queries takes approximately 66 hours to run which is a significant effort given that it has to be repeated for every new database. 
Another interesting comparison is to use the training queries also to fine-tune the zero-shot models on the IMDB database; i.e., we use zero-shot models in the few-shot mode discussed in the paper. 
As we can see, few-shot cost models that are fine-tuned on the IMDB database can further improve the cost estimation accuracy of zero-shot models. It is thus beneficial to also leverage fine-tuning in case training queries for the unseen database are available.

Finally, we can see that the MSCN models are not equally accurate which is likely due to the fact that they do not consider the physical plan that was run to execute a given query. Still, all learned approaches are more accurate than the scaled optimizer in the median after only a few queries. Furthermore, we can observe that zero-shot and few-shot cost models not only outperform workload-driven models in the median but also in the tail performance, i.e., on the 95th percentile Q-error. We can observe similar effects for the maximum Q-error.

\begin{figure}
	\centering
	\includegraphics[width=0.85\columnwidth]{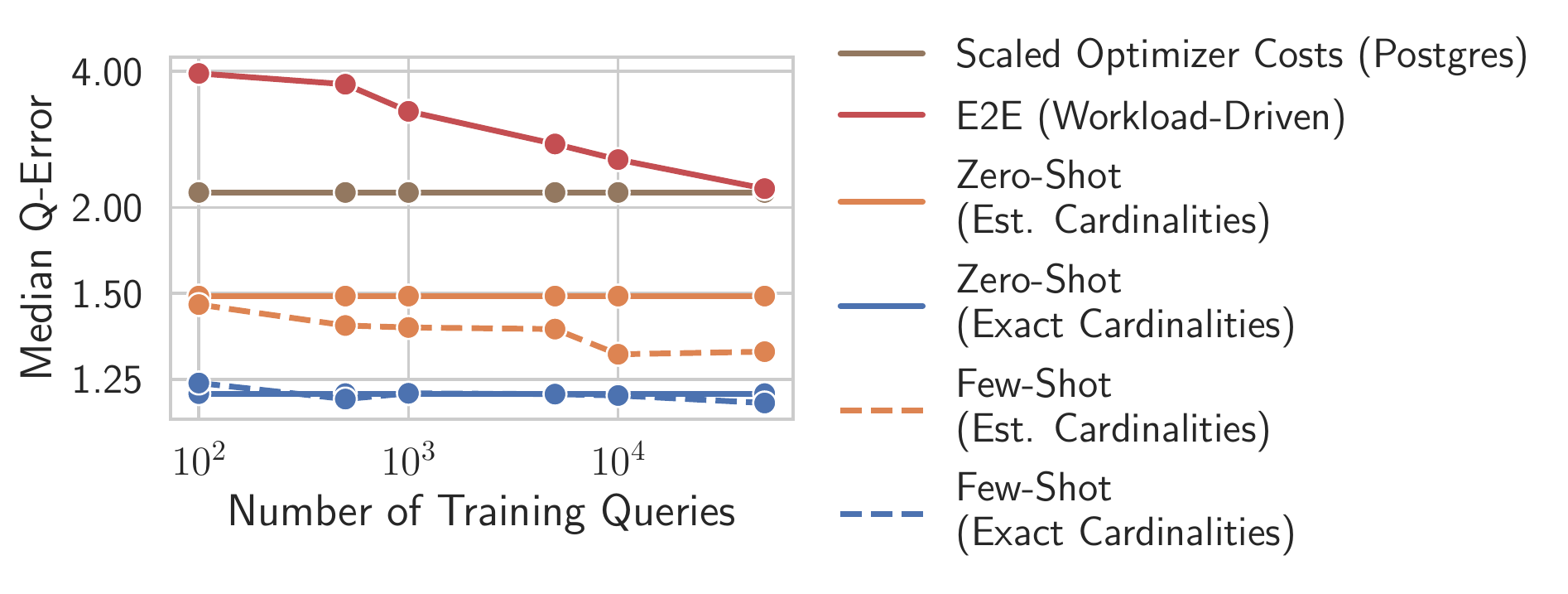}
	\vspace*{-2.5ex}
	\caption{JOB-Full Workload. Zero-shot models are significantly more accurate than the workload-driven model (E2E) and the scaled optimizer estimates even for the complex JOB benchmark. Again few-shot learning can further improve the performance of zero-shot models.}
	\vspace*{-3.5ex}
	\label{fig:job-full}
\end{figure}

\vspace*{-1.5ex}\paragraph{Complex Queries.} In this experiment, we next focus on the performance for complex queries. For this, we again train on $19$ datasets and test on the IMDB database (this time using the \textit{complex} benchmark queries) using the JOB-Full benchmark which (different from the other workloads on IMDB) contains also queries with a higher number of joins and more complex predicates including pattern-matching queries on strings. Note that data-driven models do not support complex predicates and we thus resort to the cardinality estimates of the query optimizer (Postgres) to inform the zero-shot model. As baselines, we again compare to the scaled optimizer costs and E2E which in contrast to MSCN supports complex predicates. To be fair, we use training queries with complex predicates on IMDB for the workload-driven models. In addition, we also report the accuracy of zero-shot models fine-tuned on the IMDB database using the few-shot learning.

As we can see in Figure~\ref{fig:job-full}, again zero-shot models outperform the other approaches. In particular, even the version using just optimizer cardinality estimates is more accurate than E2E using $50k$ queries which emphasizes that zero-shot cost models are robust w.r.t. imprecise cardinality estimates. The E2E models in this case need $50k$ queries just to match the performance of the scaled optimizer costs which is inferior to the previous experiment with a lower query complexity. The reason is that the E2E model has to learn the data distribution of strings as well and support complex predicates including wildcards while only observing queries. We hope that in future, data-driven models support string predicates and disjunctions as well to be used in conjunction with zero-shot cost models also for complex queries. Similar to the previous experiment, few-shot learning can further improve the accuracy.

\subsection{Exp 3: Generalization}

In this experiment, we investigate how robustly zero-shot cost models react to changes in the data characteristics and workload. 

\begin{figure}
	\centering
	\includegraphics[width=0.9\columnwidth]{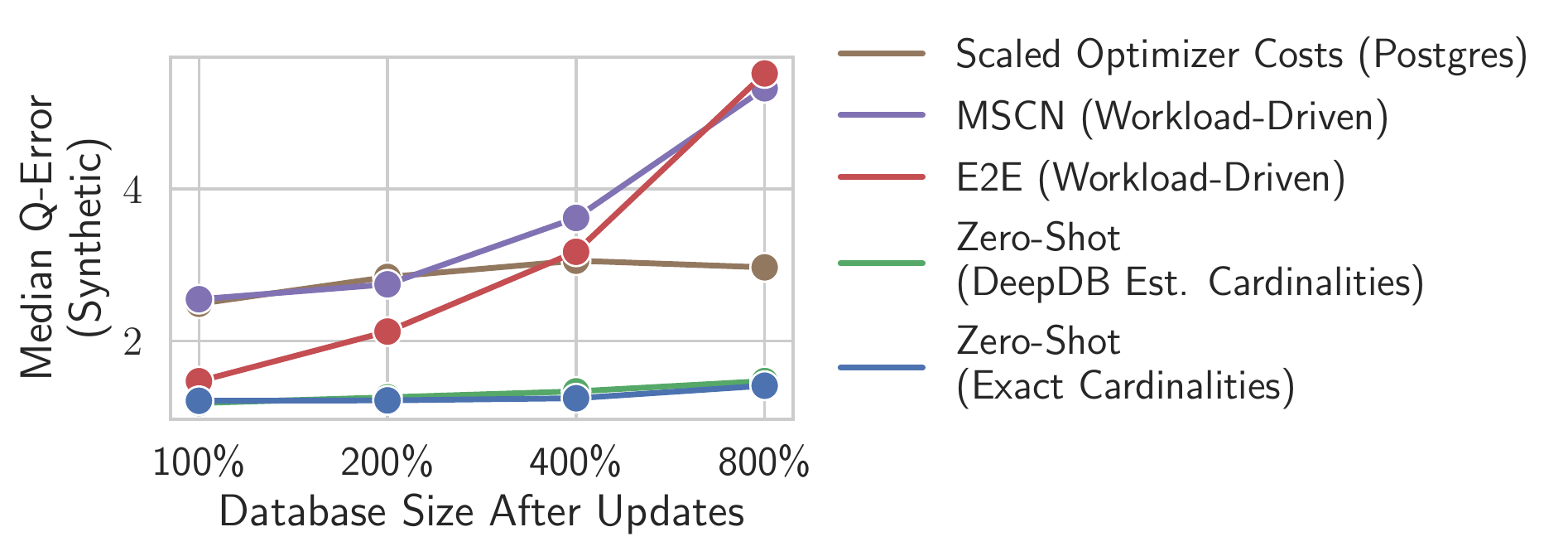}
	\vspace*{-3.5ex}
	\caption{Zero-Shot Models are robust w.r.t. Updates. Without any retraining we do not see regressions in cost estimation accuracy even for massive update rates. In contrast, workload-driven models require additional training queries.}
	\vspace*{-3.5ex}
	\label{fig:updates}
\end{figure}

\begin{figure}
	\centering
	\includegraphics[width=0.95\columnwidth]{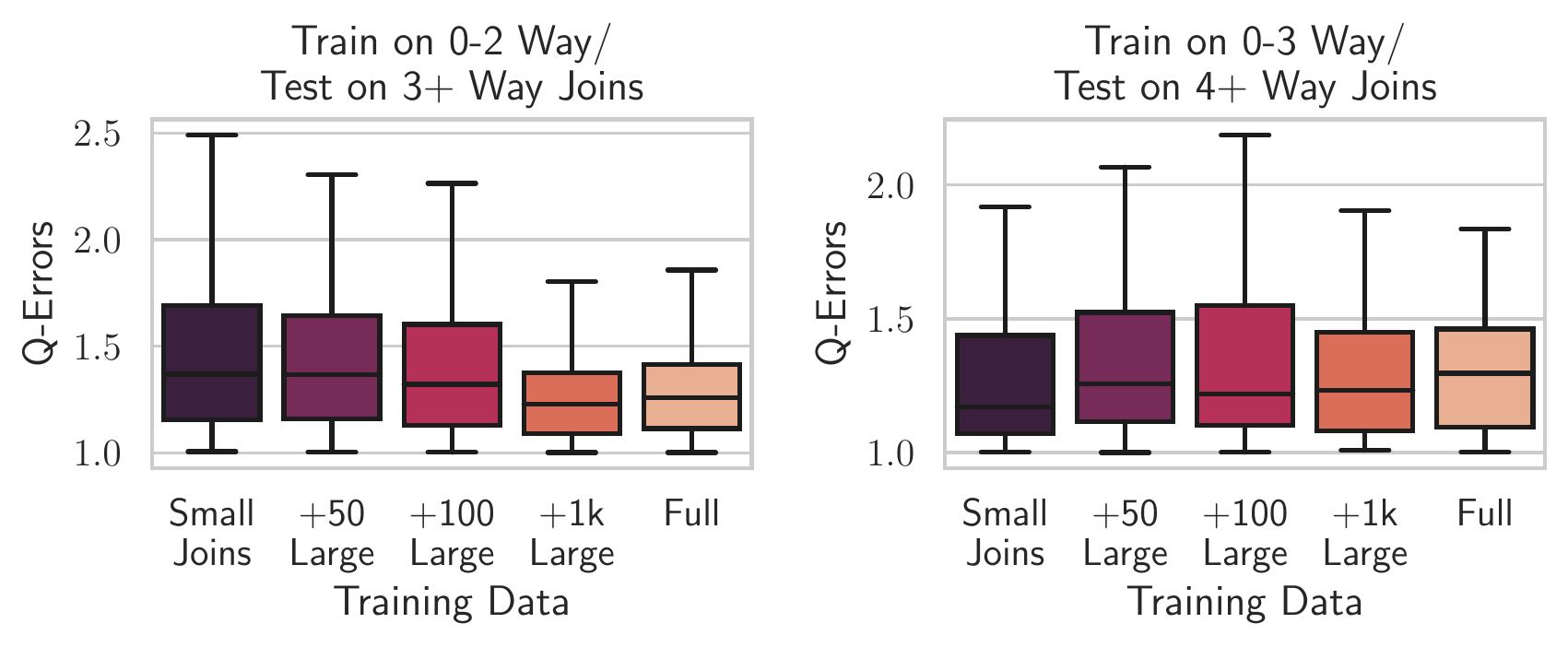}
	\vspace*{-3.5ex}
	\caption{Zero-shot cost models generalize robustly to larger Joins. Compared to zero-shot models trained also on larger joins (Full), the zero-shot models trained only on smaller joins (Small Joins) have only minor regressions in accuracy. In addition, fine-tuning the zero-shot cost models on a low number of additional queries with larger joins (resulting in few-shot models) further improves the performance.}
	\label{fig:join_gen}
	\vspace*{-3.5ex}
\end{figure}

\vspace*{-1.5ex}\paragraph{Generalization to Updates.} For the first aspect, we analyze the effects of updates on the accuracy of cost estimation.
For this, we only train on a fraction of the full data and then update the database (without retraining the prediction models). 
After the update of the database, we then predicted the query runtimes using zero-shot cost models as well as the other baselines (workload-driven models and the scaled optimizer). 
Note, that workload-driven models are expected to result in inferior performance for a higher fraction of updates since they cannot capture database updates without collecting new training data. 
This is very different from zero-shot models that get informed by data-driven models that can thus adjust to data updates without the need to retrain.
In particular, the data-driven models from DeepDB \cite{hilprecht2020deepdb} as well as classical statistics such as histograms that are compatible with zero-shot cost models are directly updateable with low overhead and hence can provide also accurate estimates after the update.

We depicted the results in Figure~\ref{fig:updates}. As we can see, there is almost no performance degradation for the zero-shot cost models with a higher update fraction. Note that we did not retrain the zero-shot cost models at all to achieve the performance but simply relied on the ability to generalize to different data characteristics. In contrast, for workload-driven models we observe a performance degradation since those models would require additional training queries on the updated database to be adapted. The reason is that the models also internalize the data distribution (i.e., table sizes and correlations) implicitly during the training and can only be informed about changes by observing additional query runtimes. This is especially problematic for more update-heavy workloads were frequently additional training queries have to be run to update the models. Note that the scaled optimizer costs do not experience such a degradation but are again less accurate than zero-shot models.

\vspace*{-1.5ex}\paragraph{Generalization to Workload Drifts.} In this experiment, we investigate how zero-shot models react to workload drifts, in particular to larger joins that appear after training a cost prediction model. To this end, we trained the zero-shot models using only queries with up to 2 or 3-way joins on the $19$ training datasets and evaluate the model using 3-way or 4-way joins (or larger) on the IMDB dataset, respectively. Since we suggest to address workload-drifts using few-shot learning, we also introduce variants that are fine-tuned on a small amount of large joins on the IMDB database.
As we can see in Figure~\ref{fig:join_gen}, the performance of the model with a  training set constrained to small joins does not degrade heavily compared to the model that was also trained on larger joins on the remaining $19$ datasets (Full) indicating a robust generalization to larger joins. 
In addition, few-shot models fine-tuned on a small amount of larger joins ($\approx50$) observed on the IMDB dataset is sufficient to achieve the same median Q-error. An even larger amount of retraining queries allows to outperform the original zero-shot model which is consistent with previous experiments showing that few-shot learning further improves the accuracy.

\begin{table}
	\scriptsize
	\centering
	\begin{tabular}{llll}
		\toprule
		& Scale & Synthetic & JOB-light \\ 
		\midrule
Scaled Optimizer Costs (Cloud DW) & 3.90 & 4.64 & 4.31 \\
Zero-Shot (DeepDB Est. Cardinalities) & 2.06 & 1.74 & 1.91 \\
Zero-Shot (Exact Cardinalities) & 1.78 & 1.56 & 1.59 \\
        \bottomrule
	\end{tabular}
	\caption{Q-errors for IMDB benchmarks on a commercial distributed cloud data warehouse.}
	\vspace{-9.5ex}
	\label{tab:cdw_q_errors}
\end{table}

\subsection{Exp 4: Extensions}

We now investigate how zero-shot models can be extended to support distributed DBMSs and different physical designs.

\vspace*{-1.5ex}\paragraph{Distributed DBMSs.} In this experiment, we executed the \textit{standard} queries on a commercial cloud DBMS and use the models to predict query runtimes of the IMDB benchmarks on this system. To this end, we adapted our zero-shot model architecture as described in Section~\ref{sec:distributed_dbms} and compare it with the scaled cost estimates of the internal query optimizer of the system. To the best of our knowledge, workload-driven approaches for cost estimation do not support distributed DBMSs as of today which is why we could not employ them as a baseline. The results are given in Table~\ref{tab:cdw_q_errors}. As we can see, zero-shot cost models are already able to outperform the cost estimates of the internal query optimizer of the DBMS. We believe that with more targeted features, the performance could even be improved further.

\vspace*{-1.5ex}\paragraph{Physical Designs.} Second, we investigate how robustly zero-shot cost models generalize to unseen physical designs - in particular using an unseen set of indexes. We again train the zero-shot cost model on $19$ databases and evaluate it using the IMDB database where this time indexes are created during the execution of the training workload. 
For training, we use \textit{index} workloads of our benchmark that involves query executions using indexes on the other $19$ databases. On the IMDB database with different indexes, we observed median Q-errors of $1.21$, $1.28$ and $1.34$ using the zero-shot variants using exact, DeepDB-estimated and cardinalities estimated by the Postgres optimizer, respectively which is comparable to the Q-errors reported before without indexes.

\begin{figure}
	\centering
	\subcaptionbox{Required Training Queries.\label{fig:req_train_queries}}[0.9\linewidth]{\includegraphics[width=0.99\linewidth]{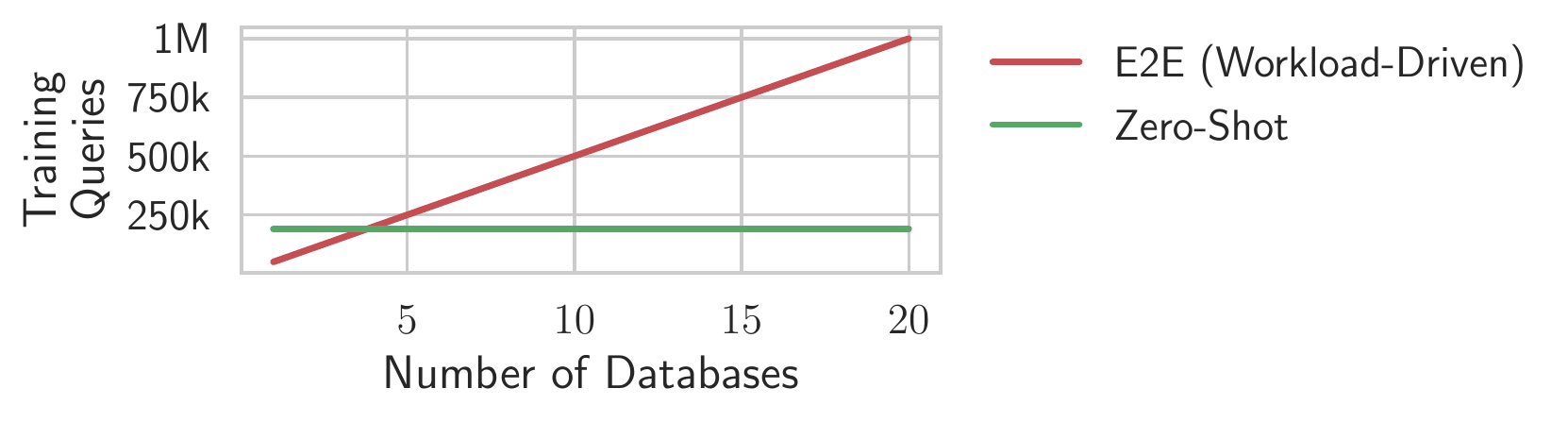}\vspace*{-1.5ex}}\\
	\subcaptionbox{Train and Test Throughput.\label{fig:train_test_throughput}}[0.9\linewidth]{\includegraphics[width=0.99\linewidth]{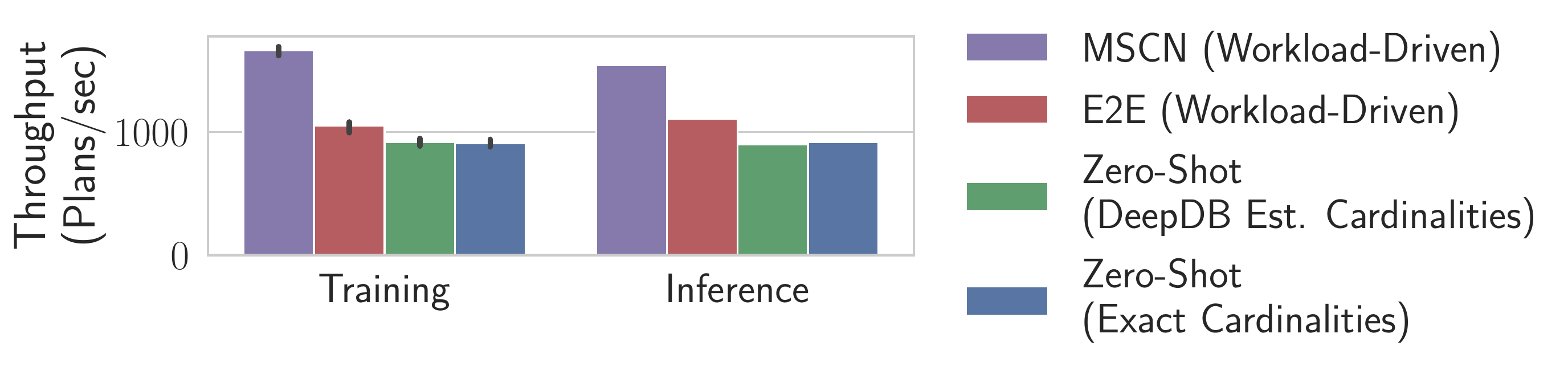}\vspace*{-1.5ex}}
	\vspace{-2.5ex}
	\caption{Training and Inference and Performance. Even though zero-shot models generalize across databases they almost match the inference and training throughput of the most accurate workload-driven alternative (E2E) and quickly amortize in terms of required training queries.
	}
	\vspace*{-2.5ex}
\end{figure}

\subsection{Exp 5: Efficiency of Training and Inference}

In this experiment, we evaluate the efficiency of training and inference of zero-shot models compared to workload-driven models.

\paragraph{Training Overhead.} In a first experiment, we compare the number of training queries required for zero-shot models as well as for workload-driven models. 
Importantly, workload-driven models need to be trained on every single database while zero-shot models can (once trained) be applied to many different databases out-of-the-box.
For showing this effect we analyze how many training queries would be required for supporting a varying number of unseen databases for which new cost estimates are required
The results are shown in Figure~\ref{fig:req_train_queries}. 
As we can see since workload execution is a one-time effort for zero-shot models (since they generalize across databases) this quickly amortizes compared to workload-driven learning since for workload-driven models, we need to collect training data for every new database.

\paragraph{Training and Inference Throughput.} In a second experiment, we compare the training and inference throughput of zero-shot cost models with state-of-the-art workload-driven approaches. 
In this experiment, we aim to show that zero-shot models are not imposing higher overhead for training and inference and thus can be used efficiently in real DBMSs. 
As we can see in Figure~\ref{fig:train_test_throughput}, zero-shot models achieve a comparable throughput and thus do not impose higher overhead comapred to workload-driven models. As we can see, the MSCN models achieve higher throughput compared to all other models (zero-shot and E2E). The reason is that these models featurize the physical query plan resulting in larger graphs compared to MSCN models which only encode the joins, tables and predicates in a query. However, this comes at the cost of an inferior predictive performance as shown before.

\begin{figure}
	\centering
	\includegraphics[width=0.9\linewidth]{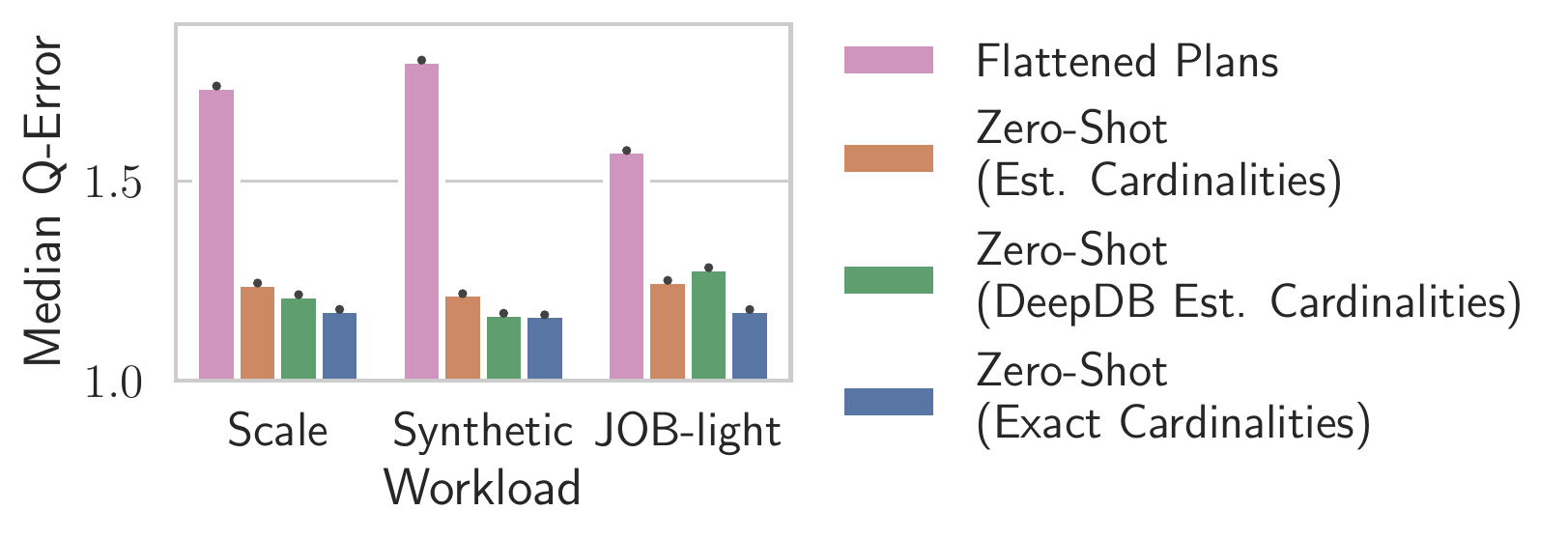}
	\vspace*{-2.5ex}
	\caption{Ablation Study. Using a flattened representation of the plans instead of our graph-based encoding yields less accurate models. Zero-shot models using the cardinality estimates of the query optimizer are still reasonably accurate.}
	\label{fig:design_space}
	\vspace*{-3.5ex}
\end{figure}

\begin{figure}
	\centering
	\includegraphics[width=0.9\columnwidth]{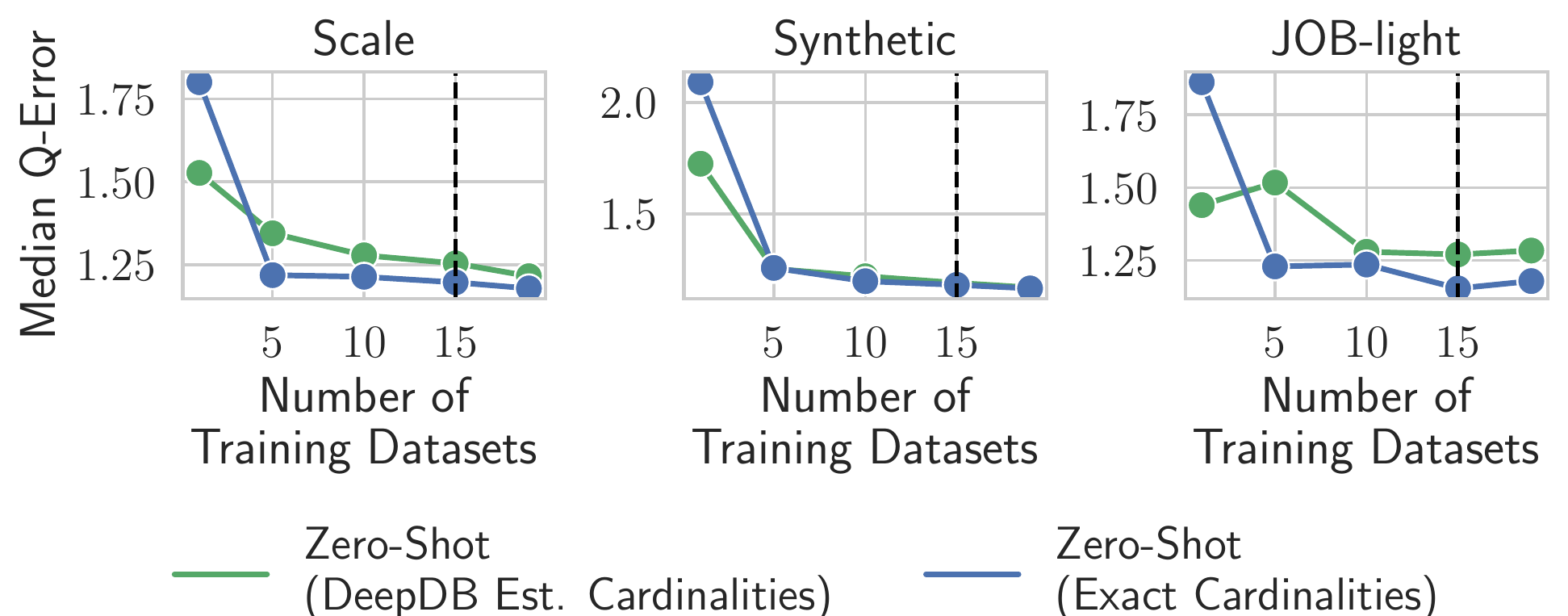}
	\vspace*{-2.5ex}
	\caption{Zero-Shot Generalization by Number of Training Databases. If we use more than 15 training databases we start to see diminishing returns in accuracy suggesting that the variety of databases in the benchmark is sufficient.}
	\label{fig:no_datasets}
	\vspace*{-2.5ex}
\end{figure}

\subsection{Exp. 6: Ablation Study}

In this experiment, we present the results of our ablation study showing the effects of the different design choices as well as the efficiency of our estimator to determine how many different databases are needed for training a zero-shot model. 

\vspace*{-1.5ex}\paragraph{Zero-Shot Design Space.} We first explore the different design space options of zero-shot cost estimation. In particular, we  focus on the questions how different cardinality estimation techniques impact the model accuracy and whether our new model architecture using graph encodings is actually required or a simpler architecture suffices. 

To address the latter question, we implemented a different version of zero-shot cost estimation that represents a single query plan as a flat vector (instead of using a graph). 
In particular, the chosen representation is similar to \citet{ganapathi2009predicting} that represents a query plan using a vector where each physical operator corresponds to two entries in the vector: one that counts how often the operator appears in the plan and one that sums up the cardinality estimates for that operator. For instance, if we only had sequential scans and nested loop joins in the query plans and one plan would scan two relations of 1M tuples each and join them resulting in 1M tuples, the vector representing the query plan could be $(2,2M,1,1M)$. Given this representation, we train a state-of-the-art regression model \cite{NIPS2017_6449f44a} to predict the runtime given a vector. Similar to the zero-shot models, we train on the remaining $19$ datasets and evaluate the performance on the IMDB benchmarks.

As we can see in Figure~\ref{fig:design_space}, the flattened version of zero-shot cost models is significantly less accurate than our proposed transferable graph-based representation. The reason is that the interactions of physical operators in the plan can only be modeled approximately if represented as a vector while our graph-based encoding allows the neural model to capture such interactions more accurately. Second, regarding cardinality estimates, we can see that data-driven cardinality estimates improve the accuracy of zero-shot cost models compared to models using optimizer cardinality estimates. However, the estimates are still very accurate even if cardinality estimates are annotated from simple cardinality estimation models that are used in DBMSs today. This is especially useful for query types that data-driven models do not support as of today and where the optimizer cardinality estimates hence serve as a fallback.

\vspace*{-1.5ex}\paragraph{Number of Training Databases.} As described in Section~\ref{sec:estimating_generalization}, in order to assess whether a zero-shot cost model has seen a sufficient number of training databases and workloads, we estimate the expected generalization error for a varying number of training databases. The generalization error is estimated by computing the test error on an unseen holdout database. If the model performance plateaus for a certain number of training databases, we can conclude that the number of training databases is sufficient.

In this experiment, we show how the generalization error develops for a growing number of training databases (i.e., from just using one up to all $19$ databases). For estimating the generalization error, we use the standard benchmark queries as defined on the IMDB dataset (i.e., we use the  synthetic, scale and JOB-light \cite{kipf2019learned} workloads).
As we can observe in Figure~\ref{fig:no_datasets}, as expected the generalization errors reduce with a growing number of databases. This is the case because with an increased number of databases the model can observe a larger variety of different data characteristics and can thus more robustly predict the runtimes for an unseen database, i.e., IMDB in this case. Interestingly, we can already achieve a reasonably small generalization error after just five different databases indicating that a moderate number of databases can be sufficient for zero-shot learning. Moreover, we clearly observe diminishing returns between $15$ to $19$ databases. 

We can thus conclude that the number of training datasets from the benchmark is indeed sufficient to allow a zero-shot model to generalize robustly to unseen databases from the benchmark and that further datasets will likely not improve the model performance. 

\vspace{-1.5ex}\section{Related Work}
\label{sec:related-work}

\paragraph{Learned Cost Estimation.}
Closest to our work are workload-driven approaches for cost estimation. 
Neural predictions models \cite{sun2019endtoend,marcus2019plan} have been proposed for cost estimation by featurizing the physical query plan as a tree. However, the models are workload-driven and thus require thousands of query executions for an unseen database. Recently, a framework has been proposed to efficiently gather this training data \cite{ventura2021expand}. In contrast, zero-shot learning completely alleviates the need to run a representative workload for new databases.
Moreover, workload-driven models were extended by improving inference and training performance \cite{kang2021efficient} and to concurrent query latency prediction \cite{zhou2020concurrentperformance}. These ideas are orthogonal and could potentially be applied to zero-shot learning as well.
An alternative to reduce the required training queries for cost estimation is DBMS fitting \cite{hilprecht2020dbmsfitting} where the idea is to model the operator complexity and adjust this basic model by fitting the parameters using differentiable programming. However, the operator complexity has to be modeled explicitly which can be impossible for complex queries. 

Earlier work proposes to use statistical methods to predict the costs of queries. For instance, it was proposed to learn models at a per-operator level \cite{mert2012learning, jiexing2012robust} to predict the overall query runtime. However, since interactions of operators cannot be learned and the models are thus too simplistic, the performance is inferior to workload-driven approaches \cite{marcus2019plan}. An alternative idea is to represent query plans as flat vectors \cite{ganapathi2009predicting} to treat cost estimation using supervised regression which we have shown to be less accurate than zero-shot cost estimation (cf. Section~\ref{sec:exp_zero_shot_eval}).

In addition, it was suggested to leverage query executions on smaller data samples or different hardware instantiations 
\cite{ferguson2012jockey, venkataraman2016ernest} or queries sharing common subexpressions \cite{siddiqui2020cost,wu2018sharedclouds} to more accurately predict the costs. In both cases, the test workload needs to closely resemble the train workload for the models to be effective again limiting the applicability. In contrast, we have shown that zero-shot models generalize to a diverse set of workloads.

\vspace{-1.5ex}\paragraph{Learned DBMS components and Design Advisors}
Machine learning has been applied more broadly to optimize DBMS systems by replacing traditional approaches for tasks such as query optimization \cite{marcus2019neo,marcus2021bao, Krishnan2018LearningTO,marcus2018deep} or query scheduling \cite{sheng2019scheduling, mao2019learning}. In addition, it was applied to knob tuning \cite{zhang2019cloudtuning}, materialized view selection \cite{han2021autonomousmv, liang2019opportunisticview}, index selection \cite{lan2020index} or partitioning \cite{hilprecht2020partitioning}. Note that all these approaches are workload-driven since query executions on the test database are required to train the models. 
We believe that zero-shot cost estimation could be used to support a variety of these tasks since they crucially depend on accurate cost estimates. 

\section{Conclusion and Future Work}
\label{sec:conclusion}

In this paper, we have demonstrated that it is possible to accurately and robustly predict query runtimes on entirely unseen databases, i.e., in a zero-shot setting. 
In addition, fine-tuning the zero-shot models to obtain few-shot models can further improve the performance if training queries are available on the new database. 
We enabled this by deriving a transferable representation of queries that generalizes across databases and a specialized model architecture. 

As a future direction, we argue that zero-shot learning has even a much broader applicability and could be applied to a large set of learned DBMS components including design advisors etc. Furthermore, we believe that the underlying principles can be applied to an even broader set of data systems (e.g., data streaming systems). 
 
\section{Acknowledgments}

This research and development project is funded by the German Federal Ministry of
Education and Research (BMBF) within the “The Future of Value Creation – Research on Production, Services and Work” program and managed by the Project
Management Agency Karlsruhe (PTKA). The author is responsible for the content of this
publication. In addition, the research was partly funded by the Hochtief project \emph{AICO} (AI in Construction), the HMWK cluster project \emph{3AI} (The Third Wave of AI), as well as the DFG Collaborative Research Center 1053 (MAKI). Finally, we want to thank the Amazon Redshift team for valuable discussions.
 

\bibliographystyle{ACM-Reference-Format}
\bibliography{bib}

\end{document}